\documentclass[twocolumn,
showpacs,
preprintnumbers,
amsmath,
amssymb,
amsfonts,
final,
a4paper,
aps,
citeautoscript,
footinbib,
pra,
superscriptaddress]{revtex4}
\usepackage{times}
\usepackage{graphicx}
\usepackage[latin1]{inputenc}

\graphicspath{{./Figs/}}
\def\Ham{\mathcal{H}}

\def\fidelity{F} 
\newcommand{\ket}[1]{\left\vert{#1}\right\rangle}
\newcommand{\bra}[1]{\left\langle{#1}\right\vert}
\newcommand{\md}[1]{\vert{#1}\vert}

\newcommand{\moy}[1]{\langle{#1}\rangle}
\newcommand{\inter}[2]{\left\langle{#1}\vert{#2}\right\rangle}
\newcommand{\elem}[3]{\left\langle{#1}\vert{#2}\vert{#3}\right\rangle}

\begin{document}

\author{Guillaume Roux}
\affiliation{LPTMS, Universit\'{e} Paris-Sud, CNRS, UMR 8626, 91405
  Orsay, France.\\
 and Institute for Theoretical Physics C, RWTH Aachen
  University, D-52056 Aachen, Germany}

\date{26 February 2009}

\title{Quenches in quantum many-body systems: One-dimensional
  Bose-Hubbard model reexamined}


\begin{abstract}
  When a quantum many-body system undergoes a quench, the
  time-averaged density-matrix $\overline{\rho}$ governs the
  time-averaged expectation value of any observable. It is therefore
  the key object to look at when comparing results with equilibrium
  predictions. We show that the weights of $\overline{\rho}$ can be
  efficiently computed with Lanczos diagonalization for relatively
  large Hilbert spaces. As an application, we investigate the
  crossover from perturbative to non-perturbative quenches in the
  nonintegrable Bose-Hubbard model: on finite systems, an approximate
  Boltzmann distribution is observed for small quenches, while for
  larger ones the distributions do not follow standard equilibrium
  predictions. Studying thermodynamical features, such as the energy
  fluctuations and the entropy, show that $\overline{\rho}$ bears a
  memory of the initial state.
\end{abstract}

\maketitle 

Recent experiments~\cite{Experiments} in ultra-cold atoms have renewed
the interest for the time-evolution of an isolated quantum many-body
system after a sudden change of the Hamiltonian parameters, the
so-called ``quantum quench''. Many questions arise from such a setup,
among which are the relaxation to equilibrium statistics, the memory
kept from the initial state, and the role of the integrability of the
Hamiltonian. Analytical and numerical results support different
answers to these
questions~\cite{Quenches,Kollath2007,Manmana2007,Kollar2008}, though
most of them have shown that observables do not follow usual
equilibrium predictions. As it has been pointed
out~\cite{Kollar2008,Relaxation}, looking at simple observables, yet
experimentally accessible, might not be considered as sufficient to
fully address these questions. Since time-evolution is unitary, there is
no relaxation in the sense of a stationary density-matrix, contrary to
what can happen in a subsystem~\cite{Relaxation}. However, observables
will fluctuate with time around some average. Standard definitions
show that the time-averaged density-matrix $\overline{\rho}$ of the
system governs any observable and its fluctuations. It is therefore
desirable to have a systematic way of getting some information about
$\overline{\rho}$, and its associated thermodynamical-like quantities, in
order to compare it with the density-matrices of equilibrium
ensembles, such as the microcanonical or the canonical ensemble.

In this paper, we show how Lanczos diagonalization (LD) enables one to
calculate the weights of the time-averaged density-matrix. This
method, which gives access to relatively large Hilbert spaces, is
helpful when an analytical calculation of the many-body wave-functions
is lacking: this is, for instance, the case of nonintegrable models. As
an application, the example of a quench in the one-dimensional
Bose-Hubbard model (BHM) is revisited for the following reasons: (i)
the model corresponds to realistic experiments~\cite{Experiments},
(ii) it is nonintegrable and it is usually believed that the
redistribution of momenta through scattering causes thermalization,
(iii) complementary numerical results already
exist~\cite{Kollath2007}, (iv) there is an equilibrium critical point
demarcating two phases, and the latter can play a role in
out-of-equilibrium physics. On finite systems, we show that there are
two distinct regimes depending on the quench amplitude: in the
perturbative regime, an approximate Boltzmann law is observed, while
distributions which do not belong to equilibrium ensembles emerge for
large quenches. Moreover, we show that the mixed state
$\overline{\rho}$ bears some memory of the initial state through its
energy fluctuations and its entropy.

We start by recalling~\cite{Peres1984} and introducing some
definitions. From now on, the discussion will be restricted to
finite-size systems of length $L$ with no accidental degeneracy. We
address the issue of the thermodynamical limit by looking at the
scaling of observables with $L$, and by giving scaling arguments for
the energy fluctuations. At time $t<0$, the Hamiltonian is denoted by
$\Ham_0$ and its eigenvectors and eigenvalues by $\ket{\psi_n}$ and
$E_n$. The system is prepared in some state $\ket{\psi_0}$, that
usually is the ground-state of $\Ham_0$. At $t=0$, the Hamiltonian is
changed to $\Ham$ which eigenvalues and eigenvectors are $\omega_n$
and $\ket{\phi_n}$. The time-evolving density-matrix of the whole
system reads $\rho(t) = \sum_{n} p_n \ket{\phi_n}\bra{\phi_n} +
\sum_{n<m} \sqrt{p_n p_m} [ e^{-i\Omega_{nm}t + i\Theta_{nm}}
  \ket{\phi_n}\bra{\phi_m} + h.c.]$, with the relative phases
$\Theta_{nm} = \theta_{n} - \theta_{m}$, using $\theta_{n} =
\text{Arg}\inter{\phi_n}{\psi_0}$, and the frequencies $\Omega_{nm} =
\omega_n - \omega_m$. The $p_n = \md{\inter{\psi_0}{\phi_n}}^2$ are
the diagonal weights of the density-matrix, and they satisfy $\sum_n
p_n = 1$. As we are generally interested in the time-averaged
expectation value of an observable $O$, we define $\overline{O} =
\lim_{t \to \infty} \frac{1}{t}\int_0^t \text{Tr} [\rho(s) O] ds =
\sum_{n} p_n O_{nn}$, with the matrix elements $O_{nm} =
\elem{\phi_n}{O}{\phi_m}$.  Interestingly, averaging $\moy{O}_0$ [with
  the notation $\moy{\cdot}_0 = \bra{\psi_0}\cdot\ket{\psi_0}$] over
random initial phase differences $\Theta_{nm}$ gives back
$\overline{O}$, relating the time-averaging to the loss of information
on the initial phases.  Similarly, by averaging $\rho(t)$ over time,
one gets
\begin{equation*}
\overline{\rho} = \sum_{n} p_n \ket{\phi_n}\bra{\phi_n}\;,
\end{equation*}
which governs any time-averaged observable since $\overline{O} =
\text{Tr} [\overline{\rho} O]$. Furthermore, it has been very recently
shown~\cite{Work} that $\overline{\rho}$ is the experimentally
relevant object to look at, and that the $p_n$ weights enter in the
microscopic expression of the work and heat done on the system in the
quench. Notice that the evolving state is a pure state so its von
Neumann entropy $S[\rho] = -\text{Tr}[\rho \ln \rho]$ is zero, while
$S[\overline{\rho}]$ is non-zero due to the loss of information
induced by time-averaging. In addition, one must also look at the
time-averaged fluctuations $\Delta O = [\text{Tr}[\overline{\rho} (O -
\overline{O})^2]]^{1/2}$ of the observables. We finally mention that,
if $O$ is diagonal in the $\ket{\phi_n}$ basis, like the energy
$\Ham$, the time-averaged expectations and fluctuations are fixed by
the initial state: $\overline{O} = \moy{O}_0$ and $\Delta O =
[\moy{(O-\overline{O})^2}_0]^{1/2}$.

The difficulty for a given system is to compute the weights $p_n$ or
any expectation value. When there is no analytical approach, as for
the BHM, a possible solution is to resort to numerical techniques.  In
order to compute the $p_n$, we notice that they enter in the
expression of the (squared) fidelity~\cite{Peres1984} $\fidelity(t)=
\md{A(t)}^2 = 1 - 4\sum_{n<m} p_n p_m \sin^2[\Omega_{nm}t/2]$. This is
the revival probability after a time $t$ because we have $A(t) =
\inter{\psi(t)}{\psi_0}$, with $\ket{\psi(t)}$ the time-evolving
wave-function. A direct time-evolution calculation usually fails after
some time~\cite{Kollath2007}. Our idea is to use spectral
methods~\cite{Dagotto1994,Mila1996} to get the Fourier transform
$A(\omega)$ of the $A(t)$ function. Contrary to the approach of
Ref.~\onlinecite{Mila1996}, we notice that LD also gives a direct
access to the Lehmann representation $A(\omega) = \sum_{n} p_n
\delta(\omega - \omega_n + E_0)$ without a finite broadening, which
induces an artificial decay of $A(t)$. Hence, all the information we
need to discuss the statistical features of $\overline{\rho}$ is
included in $A(\omega)$, since both the energies and the weights are
obtained. LD is not an exact method but is well adapted to
low-energies, i.e. long times, and we give below a perturbative
argument corroborating that the $p_n$ have an overall decrease with
$\omega_n$ [see also~\cite{EPAPS} for cross-checking]. Hilbert spaces
of sizes up to $10^7$ states will be studied in the following while
our full diagonalizations are restricted to $5000$ states. Lastly,
spectral methods being much faster than time-evolution ones, one can
scan a wide range of parameters.

The short and long time behaviors of $\fidelity(t)$ also contain
information about the $p_n$ distribution~\cite{Peres1984}: at short
times $\fidelity(t) \simeq 1 - t^2/\tau^2$ with $\tau^{-1} = \Delta
E$, the energy fluctuations. Physically, the typical time $\tau$ is
the time after which the system has ``escaped'' from the initial
state, and is the inverse of the centered width of $A(\omega)$. More
generally, higher moments of the $A(\omega)$ function are defined by
$M_q = \moy{[\Ham - \moy{\Ham}_0]^q}_0$, and are clearly fixed by the
initial state. In practice, the moments can also be independently
computed with LD for $q$ up to hundred by iteratively applying $\Ham$
on $\ket{\psi_0}$. The associated sum rules are useful to cross-check
the calculation of the spectrum. If one understands $A(\omega)$ as a
probability distribution, knowing all moments amounts to knowing the
distribution itself and would give back the exact $\overline{\rho}$.
This comment was put forward without proof in
Ref.~\onlinecite{Manmana2007}, together with a relevant discussion on
the relation between these moments and generalized Gibbs ensembles. At
long times, $\fidelity(t)$ usually fluctuates around its mean value
$\bar{\fidelity} = \sum_n p_n^2$~\cite{Peres1984}. A qualitative
interpretation of $\bar{\fidelity}$ is the ``participation
ratio''~\cite{Peres1984} that counts the number of eigenstates which
contributes to time evolution. The typical fluctuations of the
fidelity are $(\Delta \fidelity)^2 = \overline{\fidelity(t)^2 -
  \bar{\fidelity}^2} = 4 \sum_{n<m} p_n^2 p_m^2$. This quantity
measures the strength of the wavering of the evolving state between
getting back to $\ket{\psi_0}$ or getting away from $\ket{\psi_0}$.

Qualitatively, a quench consists in projecting the initial state onto
the spectrum of the Hamiltonian $\Ham$ governing the dynamics.
Straightforward results from perturbation theory in the quench
amplitude illustrate the difference between small and large quenches:
one expects a crossover between the two regimes.  Writing $\Ham =
\Ham_0 + \lambda \Ham_1$ with $\lambda$ the quench amplitude and
$\Ham_1$ the perturbing operator, the perturbed weights read, for
$\lambda \ll 1$, $ p_0 \simeq 1 - \lambda^2 \sum_{n \neq 0} h_{n0}$,
and $p_{n\neq 0} \simeq \lambda^2 h_{n0}$, in which the notation
$h_{n0} = \md{\elem{\psi_n}{\Ham_1}{\psi_0} }^2 / (E_n-E_0)^2$ has
been used. Meanwhile, the $\omega_n$ are slightly shifted to order
$\lambda$ and the eigenfunctions too. Thus, the $p_n$ have an overall
decrease with the excited energy and, increasing $\lambda$ induces a
transfer of spectral weight from the ``targeted'' ground-state
$\ket{\phi_0}$ to other excited states. We get the scaling of several
quantities to lowest order in $\lambda$: $M_q \propto \lambda^{2}$, $1
- \bar{\fidelity} \propto \lambda^2$ and $\Delta \fidelity \propto
\lambda^2$. As $\bar{\fidelity} > 0$, these scalings will naturally
fail for large $\lambda$, signaling the crossover to the
non-perturbative regime. In addition, we mention that the mean-energy
$\moy{E}$ is simply always linear in $\lambda$, since we have $\moy{E}
= \moy{\Ham}_0 = E_0 + \lambda \moy{\Ham_1}_0$. 

\begin{figure}[b]
\includegraphics[width=0.9\columnwidth,clip]{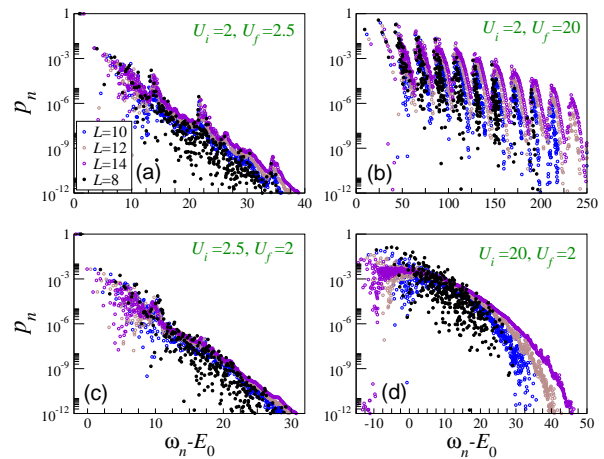}
\caption{(Color online) Distributions of the $p_n$ at four different
  points of the $(U_i,U_f)$ state diagram. For the smallest size
  $L=8$, \emph{exact} results are obtained by full diagonalization.}
\label{fig:distributions}
\end{figure}

\begin{figure*}[t]
\includegraphics[height=0.22\textwidth,clip]{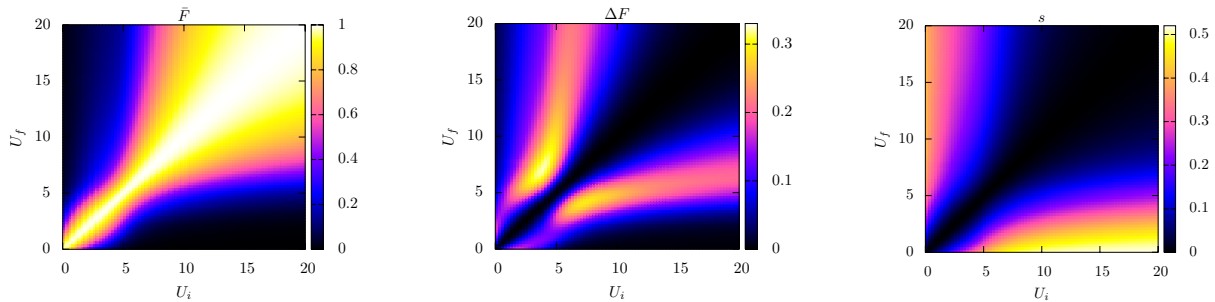}
\caption{(Color online) Maps of the observables $\bar{\fidelity}$,
  $\Delta \fidelity$ and entropy per particle $s$ characterizing the
  time-averaged density-matrix $\overline{\rho}$. Results are obtained
  by LD on a finite system ($L=12$) with periodic boundary
  conditions.}
\label{fig:maps}
\end{figure*}

\textit{Application to a quench in the one-dimensional Bose-Hubbard
  model} -- We now study the BHM in a one-dimensional optical lattice
which is a nonintegrable model:
\begin{equation*}
\Ham = -J \sum_j [b^{\dag}_{j+1} b_j + b^{\dag}_{j} b_{j+1} ] 
        + U/2 \sum_j n_j(n_j-1)\,,
\end{equation*}
with $b^{\dag}_j$ the operator creating a boson at site $j$ and $n_j =
b^{\dag}_j b_j$ the local density. $J$ is the kinetic energy scale
while $U$ is the magnitude of the onsite repulsion. In an optical
lattice, the ratio $U/J$ can be tuned by changing the depth of the
lattice and using Feshbach resonance~\cite{Experiments}. When the
density of bosons is fixed at $n=1$ and $U$ is increased, the
equilibrium phase diagram of the model displays a quantum phase
transition from a superfluid phase to a Mott insulating phase in which
particles are localized on each site. The critical point has been
located at $U_c \simeq 3.3 J$ using numerics~\cite{Kuhner2000}. The
quenches are performed by changing the interaction parameter $U_i
\rightarrow U_f$ (we set $J=1$ in the following), so we have $\lambda
= (U_f - U_i)/2$, and the perturbing operator $\Ham_1 = \sum_j n_j
(n_j-1)$ is diagonal. Numerically, one must fix a maximum onsite
occupancy. We take four as in Ref.~\onlinecite{Kollath2007} (for
further details, see~\cite{EPAPS}).

Since $\overline{\rho}$ features a mixed state, we call the
$(U_i,U_f)$ plane a state diagram. The $U_i = U_f$ ($\lambda = 0$)
line splits this state diagram in two regions and the previous
perturbative arguments should hold close to this line. The typical
distributions of the weights versus energy for four points of the
state diagram are given in Fig.~\ref{fig:distributions}: two (a,c)
with small quenches with parameters of the same (superfluid)
equilibrium phase, and two (b,d) with large quenches, in which $U_f$
``crosses'' $U_c$ in both directions. We observe that in the first two
situations, for small $\lambda$, the distributions are close to an
exponential decay typical of a \emph{canonical} ensemble. This result
supports the evidence of a ``thermalized'' regime as found in
Ref.~\onlinecite{Kollath2007}, but on more general grounds since we
directly have the distribution. Secondary peaks in
Fig.~\ref{fig:distributions}(a) yield correction to this Boltzmann
law. By looking at the cases of large quenches, we see that the
distributions are strongly different from either the microcanonical or
the canonical ensemble. When $U_f=20$
[Fig.~\ref{fig:distributions}(b)], Mott excitations, corresponding to
doubly occupied sites and roughly separated by $U_f$, are clearly
visible in the spectrum. Although the overall decay of the $p_n$ is
exponential, the distribution is very different from a Boltzmann law.
This explains that many observables differ from the ones of an
equilibrium system, and independently corroborates results of
Ref.~\onlinecite{Kollath2007}. When $U_f=2$
[Fig.~\ref{fig:distributions}(d)], the targeted spectrum is nearly
continuous and the distribution displays large weights around zero
energy and a subexponential-like behavior [approximately
$\exp(-(\omega_n - E_0)^{\gamma})$ with $\gamma > 1$]. This is again
different from equilibrium predictions. The bump-like shape of the
$U_f=2$ distribution can be qualitatively understood from the fact
that the ground-state energy increases with $U$ in the BHM. As $E_0 >
\omega_0$ when $U_f < U_i$, the initial state is close in energy to
some excited states of $\Ham$ and, according to the perturbative form
of the $p_n$, this favors their excitations by the quenching process.
Another consequence is that the state diagram is expected to be
non-symmetrical with respect to the $U_i=U_f$ line.

\begin{figure}[t]
\includegraphics[width=0.9\columnwidth,clip]{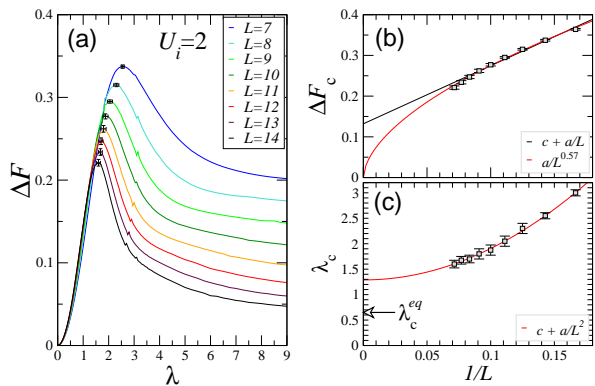}
\caption{(Color online) (a) Cut along the $U_i = 2$ line showing a
  maximum of $\Delta{\fidelity}$ between the perturbative and
  non-perturbative regimes of the quench. (b-c) Finite size scalings
  of $\Delta{\fidelity}_c$ and $\lambda_c$. See text for discussion.}
\label{fig:cut}
\end{figure}

\begin{figure}[t]
\includegraphics[width=0.7\columnwidth,clip]{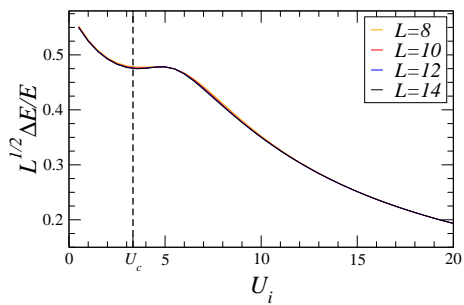}
\caption{(Color online) Rescaled relative energy fluctuations in the
  mixed state. They only depend on the features of the initial state.
  The slope of the curves vanishes close to the equilibrium critical
  point $U_c$.}
\label{fig:fluctuations}
\end{figure}

\emph{Crossover and finite size effects} -- To sketch the state
diagram, maps of integrated quantities such as $\bar{\fidelity}$,
$\Delta F$, and the entropy per particle $s=S[\overline{\rho}]/N$ are
computed on a finite system with $L=12$ and given in
Fig.~\ref{fig:maps}. The normalization of the entropy
$S[\overline{\rho}]$ is motivated by the fact that we observe that it
scales as $N$, plus some finite-size corrections. As suggested
previously, observables display a crossover from the perturbative
regime to a non-perturbative regime characterized by a significant
enhancement of the weights of excited states. In order to evaluate the
finite size effects on the crossover, we look at the scalings of
$\bar{\fidelity}$ and $\Delta F$ for a cut along the $U_i = 2$ line
and increasing $\lambda$. A first question is how the size of the
perturbating regime evolves when increasing the length $L$. To address
this question, we look at the evolution of two demarcating points.
One is associated with $\bar{\fidelity}$ and is inconclusive (for
further details, see~\cite{EPAPS}). More interestingly, $\Delta F$
scales as $\lambda^2$ in the perturbative regime and the slope
increases with $L$ (see Fig.~\ref{fig:cut}). At large $\lambda$,
$\Delta F$ is nearly flat and rapidly decreases with $L$. In between,
it passes through a maximum that defines a demarcating point
$\lambda_c(L)$, and the corresponding $\Delta{\fidelity}_c =
\Delta{\fidelity}(\lambda_c(L))$. The scaling of $\lambda_c(L)$
suggests a finite value in the thermodynamical limit [see
Fig.~\ref{fig:cut}(c)]. $\Delta{\fidelity}_c$ can scale to a finite
value but also to zero as a power-law [see Fig.~\ref{fig:cut}(b)].  We
notice that the latter situation would be in contradiction with a
finite $\lambda_c$ and the fact that $\Delta F$ increases with $L$ at
low $\lambda$. These results \emph{suggest} that the perturbative
regime survives in the thermodynamical limit, but they remain
questionable. From the extrapolations, we find that the crossover
survives for larger sizes, and could be experimentally relevant since
experiments deal with finite systems. Notice that some of the numerics
in Ref.~\onlinecite{Kollath2007} were done on larger systems.  Another
question one can ask is the role of the critical point on the observed
maximum of the fluctuations of the fidelity: one may define the
``equilibrium expectation'' $\lambda_c^{eq} = (U_c - U_i)/2$ [resp.
$(U_f - U_c)/2$] if one scans over $U_f$ [resp. $U_i$] and compare it
with the scalings of actual $\lambda_c$. In Fig.~\ref{fig:cut}(c), the
two are too close to be conclusive but for large
$U_{i,f}$~\cite{EPAPS}, the difference is much substantial and
$\lambda_c(L)$ even scales away from $\lambda_c^{eq}$. Thus, we infer
that $U_c$ certainly plays a role (see below and
Fig.~\ref{fig:fluctuations}), but not on the crossover nor on the
location of $\Delta{\fidelity}_c$.

We now discuss some of the thermodynamical features of the mixed
states described by the density-matrix $\overline{\rho}$. Firstly, we
ask whether the averaged energy is well-defined by looking at the
relative energy fluctuations defined as $\Delta E / E \equiv \Delta
E/(\moy{E}-E_0 + \lambda N) = \sqrt{\sum_{ij} \moy{n_i^2 n_j^2}_0
  -\moy{n_i^2}_0 \moy{n_j^2}_0} / \sum_{i} \moy{n_i^2}_0$ to get rid
of the obvious dependencies of $\moy{E}$ on $E_0$, $N$ and $\lambda$:
what remains are the relative ``squared density'' fluctuations in the
initial state. In the superfluid phase, we expect~\cite{Giamarchi2004}
the squared density-density correlations to have an asymptotic
algebraic behavior $\moy{n^2_i n^2_j}_0 - \moy{n^2_i}_0 \moy{n^2_j}_0
\sim |i-j|^{-\alpha}$, while they should be exponential in the Mott
phase $e^{-|i-j|/\xi}$, with $\xi$ the correlation length.  On a chain
of length $L$, we thus have $\Delta E = \lambda \sqrt{L} g(L)$ with:
(i) if $\alpha < 1$, then $g(L) \sim L^{(1-\alpha)/2}$, (ii) if
$\alpha = 1$, $g(L) \sim \sqrt{\ln L}$ and if $\alpha > 1$ or $\xi > 0
$, $g(L) = \text{const}$. As we have $\alpha > 1$ in the superfluid
phase of the 1D BHM~\cite{Giamarchi2004} and $\sum_i \moy{n_i^2}_0
\sim L$, we find that $\Delta E / E = f(U_i,L) / \sqrt{L}$ for any
$U_i$. The $f(U_i,L)$ function is computed with LD and plotted in
Fig.~\ref{fig:fluctuations}. It shows a very good agreement with this
scaling argument since $f$ hardly depends on $L$. This $1/\sqrt{L}$
scaling resembles the ones of the (micro)canonical ensembles but, one
also notices that starting from a initial state with strong density
fluctuations ($\alpha \leq 1$) leads to anomalous scalings of the
relative energy fluctuations. In the BHM, this could be achieved by
introducing nearest neighbor repulsion~\cite{Kuhner2000}. We also get
the scaling of the typical time $\tau \sim \lambda^{-1} L^{-1/2}$.
This shows that, even if $\tau$ scales to zero in the thermodynamical
limit, it can be significantly long on large but finite systems for
small $\lambda$. More importantly, we find that two mixed states
$\overline{\rho}$ can have the same energy $\moy{E}$ but with
different $\lambda$, since $\moy{E} = E_0 + \lambda \moy{\Ham_1}_0$.
Hence, each of them originates from a different initial state and
consequently, the two states have different energy fluctuations.
Consequently, $\overline{\rho}$ keeps a memory on the initial state.
Another important thermodynamical feature is the entropy per particle
$s$ that continuously increases with $\lambda$ and reveals more
significantly the underlying anisotropy of the state diagram [see
Fig.~\ref{fig:maps}]. We have checked that two mixed states with the
same mean energy $\moy{E}$ have different entropies, so
$\overline{\rho}$ also keeps a memory of the initial state through its
entropy. From the very definition of the $p_n$, this is not
surprising.

In conclusion, we have shown that the weights of the time-averaged
density-matrix $\overline{\rho}$ can be obtained with LD. This
provides an observable-free description of the quench process, in
particular for nonintegrable models. The method is applied to the 1D
BHM where it is shown that, on finite systems, there is a clear
crossover from a perturbative regime, in which the distribution is
Boltzmann-like in the superfluid region, to distributions that are not
predicted by equilibrium statistics ensembles. The state diagram has
been mapped out in the $(U_i,U_f)$ plane and finite size effects have
been investigated.  Lastly, we showed that the mixed state
$\overline{\rho}$ has a well-defined energy and that it keeps a memory
of the initial state through its energy fluctuations or its entropy.

I thank T. Barthel, F. Heidrich-Meisner, T. Jolicoeur, D. Poilblanc
and D. Ullmo for fruitful discussions.

\appendix

\onecolumngrid

\section*{{\Large Electronic physics auxiliary publication service for:
On quenches in quantum many-body systems: the
  one-dimensional Bose-Hubbard model revisited}}

\section*{Typical behavior of the fidelity with time}

\begin{figure}[!ht]
\includegraphics[width=0.4\columnwidth,clip]{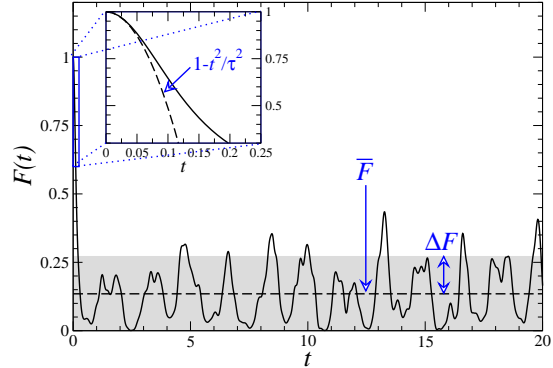}
\caption{Typical behavior of the fidelity for a finite size system
  with $L = 10$ starting from $U_i=2$ to $U_f=8$ ($\lambda =3$). At
  short times: $F(t) = 1 - t^2/\tau^2$ (here $\tau = 0.14$). For long
  times, $F(t)$ fluctuates around its mean value $\bar F$ (here
  $\bar{F} = 0.135$ and $\Delta F = 0.136$). }
\label{fig:typical}
\end{figure}

\section*{Technical details on Lanczos calculations}

\begin{figure}[!ht]
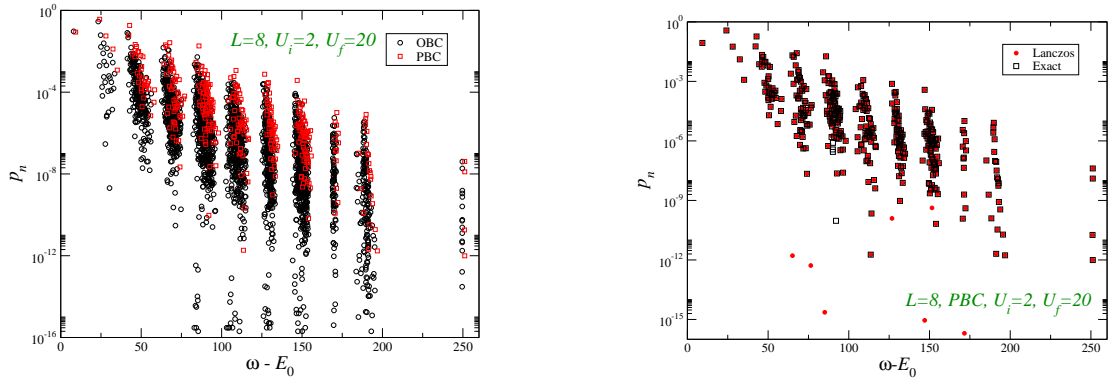

\includegraphics[width=0.35\columnwidth,clip]{fig6a}\hspace{2cm}
\includegraphics[width=0.35\columnwidth,clip]{fig6b}
\caption{Test on symmetries and effect of boundary conditions on the
  distribution of the $p_n$. PBC stands for periodic boundary
  conditions while OBC is for open BC.}
\label{fig:TestBoundaries}
\end{figure}

We use 200 Lanczos iterations to get the ground state and 1200 to get
the Lehmann representation of $A(\omega)$. We do not use symmetries of
the Hamiltonian except particle number conservation. With periodic
boundary conditions, translational symmetries induce some selection
rules for the $p_n$ so their number is quite reduced. We have checked
that Lanczos gives a good result by comparing it with exact results
obtained by full diagonalization on a system with $L=8$ (see
Fig.~\ref{fig:TestBoundaries}.  The largest Hilbert space size is
13311000 for $L=14$ for Lanczos diagonalization and 5475 for $L=8$ for
full diagonalization.  Very similar results are obtained from systems
with open boundary conditions.

\section*{Additional results on the Bose-Hubbard model}

\begin{figure}[!ht]
\includegraphics[width=0.3\columnwidth,clip]{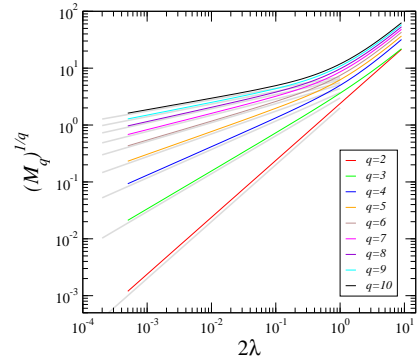}
\caption{First moments $M_q$ (to the power $1/q$) of the $A(\omega)$
  function for a system with $L=10$ and $U_i=2$. There is a crossover
  from the perturbative result $M_q \sim \lambda^{2}$ (gray lines) to
  a regime where $M_q \sim \lambda^{q}$ at large $\lambda$.}
\label{fig:moments}
\end{figure}

Moments are related to the $p_n$ and $\omega_n$ through $ M_q = \sum_n
p_n [\omega_n - \moy{E}]^q$. They undergo a clear change of behavior
with increasing $\lambda$ as shown in Fig.~\ref{fig:moments}.

\begin{figure}[!ht]
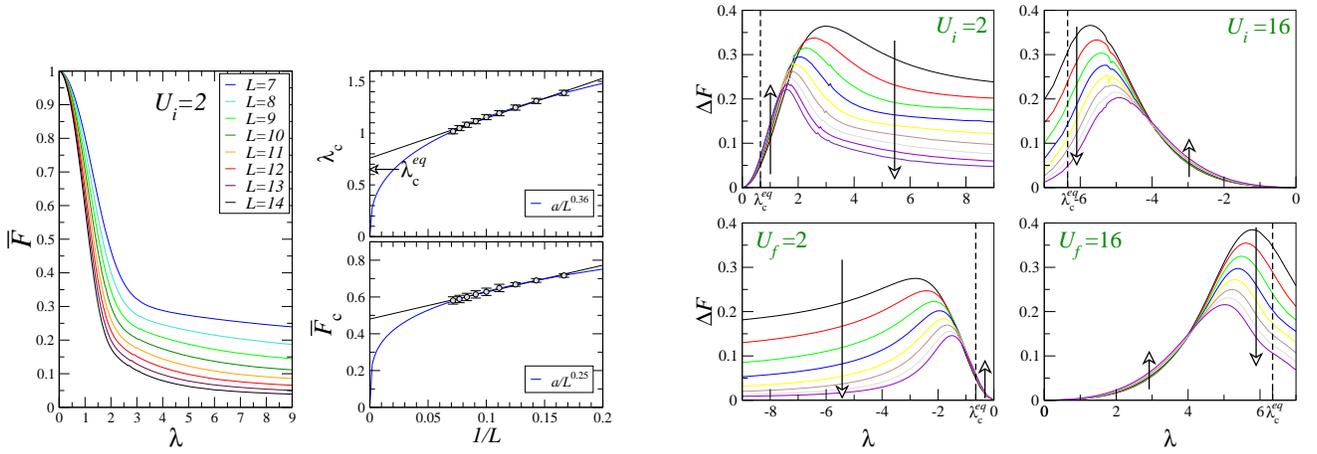

\includegraphics[width=0.45\columnwidth,clip]{fig8}\hspace{1cm}
\includegraphics[width=0.45\columnwidth,clip]{fig9}
\caption{\emph{Left:} Cut along the $U_i = 2$ axis of
  $\bar{\fidelity}$. A linear scaling gives both a finite value for
  $\bar{\fidelity}_c$ and $\lambda_c$ but a power-law one (going to
  zero) is also plausible. \emph{Right:} Four cuts in the state
  diagram showing the scaling behavior of $\Delta F(L)$ as a function
  of $\lambda$. The smallest size is $L=6$ and larger $L=13$ except
  for $U_i=2$ for which it is $L=14$. Results for $L=6,7,8$ are
  \emph{exact} (full diagonalization) while larger sizes are obtained
  with Lanczos. The arrows indicate how $\Delta F$ increases or
  decreases with $L$.}
\label{fig:cut_deltaP}
\end{figure}

We give below the behavior of $\bar{\fidelity}$ which goes from 1 when
$\lambda=0$ to small value when $\lambda$ is large. On a finite
system, the second derivative $d^2\bar{\fidelity}/d\lambda^2$ crosses
zero for a value $\lambda_c(L)$ and we define the corresponding
$\bar{\fidelity}_c = \bar{\fidelity}(\lambda_c(L))$. The scalings with
$1/L$ of these two quantities are given in Fig.~\ref{fig:cut_deltaP}:
a linear scaling suggests that they are finite in the thermodynamical
limit but power-law scalings going to zero also works for both, so
studying this quantity is not very conclusive. Power-law scalings are
however very slow, which means that even for large systems of length
100 or 1000 (experimentally relevant), the perturbative regime should
survive.

Notice that in the three other cuts of $\Delta F$
[Fig.~\ref{fig:cut_deltaP}] support that the same increase of $\Delta
F$ with $L$ in the perturbative regime as for the $U_i=2$ case of the
paper.

\end{document}